\RequirePackage[mathlines]{lineno} 
\documentclass[prd,twocolumn,showpacs,amsmath,amssymb]{revtex4}
\usepackage{overpic,graphicx}
\usepackage{dcolumn}
\usepackage{bm}
\usepackage{rotating}
\usepackage{subfigure}
\usepackage{color}
\usepackage{lineno}
\let\oldequation\equation
\let\oldendequation\endequation

\renewenvironment{equation}
  {\linenomathNonumbers\oldequation}
  {\oldendequation\endlinenomath}

\begin{document}

\title{\bf \boldmath
Observation of $D^+\to\eta\eta\pi^+$ and improved measurement of $D^{0(+)}\to\eta\pi^+\pi^{-(0)}$
}

\author{
M.~Ablikim$^{1}$, M.~N.~Achasov$^{10,e}$, P.~Adlarson$^{63}$, S. ~Ahmed$^{15}$, M.~Albrecht$^{4}$, M.~Alekseev$^{62A,62C}$, A.~Amoroso$^{62A,62C}$, Q.~An$^{59,47}$, ~Anita$^{21}$, Y.~Bai$^{46}$
% ...
\\
\vspace{0.2cm}
(BESIII Collaboration)\\
\vspace{0.2cm} {\it
$^{1}$ Institute of High Energy Physics, Beijing 100049, People's Republic of China\\
% ...
$^{4}$ Bochum Ruhr-University, D-44780 Bochum, Germany\\
% ...
$^{10}$ G.I. Budker Institute of Nuclear Physics SB RAS (BINP), Novosibirsk 630090, Russia\\
% ...
$^{15}$ Helmholtz Institute Mainz, Johann-Joachim-Becher-Weg 45, D-55099 Mainz, Germany\\
% ...
$^{21}$ Indian Institute of Technology Madras, Chennai 600036, India\\
% ...
$^{46}$ Southeast University, Nanjing 211100, People's Republic of China\\
$^{47}$ State Key Laboratory of Particle Detection and Electronics, Beijing 100049, Hefei 230026, People's Republic of China\\
% ...
$^{59}$ University of Science and Technology of China, Hefei 230026, People's Republic of China\\
% ...
$^{62}$ (A)University of Turin, I-10125 Turin, Italy; (B)University of Eastern Piedmont, I-15121 Alessandria, Italy; (C)INFN, I-10125 Turin, Italy\\
$^{63}$ Uppsala University, Box 516, SE-75120 Uppsala, Sweden\\
% ...
$^{e}$ Also at the Novosibirsk State University, Novosibirsk, 630090, Russia\\
% ...
  }
}

\begin{abstract}
Using an $e^+e^-$ annihilation data sample corresponding to an integrated luminosity of $2.93\,\rm fb^{-1}$ collected
at the center-of-mass energy of 3.773\,GeV with the BESIII detector, we measure
the absolute branching fractions of
$D^+\to\eta\eta\pi^+$, $D^+\to\eta\pi^+\pi^0$, and $D^0\to\eta\pi^+\pi^-$
to be
$(2.96 \pm 0.24 \pm 0.10)\times 10^{-3}$, $(2.23 \pm 0.15 \pm 0.10)\times 10^{-3}$, and
$(1.20 \pm 0.07 \pm 0.04)\times 10^{-3}$, respectively, where the first uncertainties are statistical and the second ones are systematic.
The $D^+\to\eta\eta\pi^+$ decay is observed for the first time,
and the branching fractions of $D^{+(0)}\to\eta\pi^+\pi^{0(-)}$ are measured with much improved precision.
In addition we test for $CP$ asymmetries in the separated charge-conjugate
branching fractions; no evidence of $CP$ violation is found.

\end{abstract}


\maketitle

\section{Introduction}

The goal of the experimental studies of hadronic $D$ meson decays is to explore strong and weak interaction effects. Various experiments have measured the
branching fractions (BFs) of hadronic decays of $D$ mesons~\cite{pdg2018}. However, measurements of singly Cabibbo-suppressed decays to final states containing one or more $\eta$ mesons are still limited~\cite{pdg2018}. Recently, the BESIII Collaboration presented measurements of $D^0\to \eta\pi^0\pi^0$ and $D^0\to \eta\eta\pi^0$~\cite{pietaeta}. The isospin-related decay modes $D^+\to\eta\pi^+\pi^0$ and $D^0\to \eta\pi^+\pi^-$ were measured with large uncertainties by the CLEO Collaboration~\cite{cleoeta}, and there is no measurement for $D^+\to \eta\eta\pi^+$. Improved measurements of $D^+\to\eta\pi^+\pi^0$, $D^0\to \eta\pi^+\pi^-$ and the search for $D^+\to \eta\eta\pi^+$ will be useful to clarify the gaps between the inclusive and known exclusive $D\to \eta X$ decay rates.
On the other hand, measurements of these decays provide important inputs for charm and $B$ physics. For instance, these multibody hadronic $D$ decays are crucial backgrounds in semitauonic decays of $B$ mesons; thus, precision measurements of these hadronic decays are important for the test of lepton flavor universality~\cite{lhcbnote}.

This paper presents the first measurement of the BFs of $D^+\to \eta\eta\pi^+$ and the improved measurements of $D^{+(0)}\to\eta\pi^+\pi^{0(-)}$ using an $e^+e^-$ data sample of 2.93\,fb$^{-1}$ taken at the center-of-mass energy $\sqrt s=$ 3.773~GeV~\cite{lum_bes3}. In order to search for $CP$ violation in $D$ decays~\cite{pap01,pap02}, the asymmetries of the BFs of the charge-conjugate decays, defined as $A_{CP}=\frac{{\mathcal B}(D\to f)-{\mathcal B}(\bar{D}\to {\overline{f}})}{{\mathcal B}(D\to f)+{\mathcal B}(\bar{D}\to {\overline{f}})}$, have also been measured for the first time. Throughout the paper, charge-conjugate modes are implied, except for the $A_{CP}$ measurements.

\section{BESIII detector and Monte Carlo simulation}

The BESIII detector is a magnetic spectrometer~\cite{Ablikim:2009aa} located at the
Beijing Electron Positron Collider (BEPCII)~\cite{Yu:IPAC2016-TUYA01}.
The cylindrical core of the BESIII detector consists of a helium-based multilayer drift chamber (MDC),
a plastic scintillator time-of-flight system (TOF), and a CsI(Tl) electromagnetic calorimeter (EMC),
which are all enclosed in a superconducting solenoidal magnet providing a 1.0~T magnetic field.
The solenoid is supported by an octagonal flux-return yoke with resistive plate muon chambers interleaved with steel.
The acceptance of charged particles and photons is 93\% of the $4\pi$ solid angle.
The charged-particle momentum resolution at $1~{\rm GeV}/c$ is $0.5\%$,
and the $dE/dx$ resolution is $6\%$ for the electrons from Bhabha scattering.
The EMC measures photon energies with a resolution of $2.5\%$ ($5\%$) at $1$~GeV in the barrel (end cap) region.
The time resolution of the TOF barrel part is 68~ps, while that of the end cap part is 110~ps.

Simulated samples produced with the {\sc
geant4}-based~\cite{geant4} Monte Carlo (MC) package which
includes the geometric description of the BESIII detector and the
detector response, are used to determine the detection efficiency
and to estimate the backgrounds.

The MC sample used includes production of $D\bar{D}$
pairs with consideration of quantum coherence for all neutral $D$
modes, the non-$D\bar{D}$ decays of the $\psi(3770)$, the initial state radiation (ISR)
production of the $J/\psi$ and $\psi(3686)$ states, and the
continuum processes incorporated in {\sc kkmc}~\cite{ref:kkmc}.
The simulation includes the beam
energy spread and ISR in the $e^+e^-$
annihilations modeled with the generator {\sc
kkmc}~\cite{ref:kkmc}.

The known decay modes of the $D$ mesons and the charmonium states are modeled with {\sc
evtgen}~\cite{ref:evtgen} using BFs taken from the
Particle Data Group~\cite{pdg2018}, and the remaining unknown decays
from the charmonium states with {\sc lundcharm}~\cite{ref:lundcharm}.
Final state radiation is incorporated with the {\sc photos}
package~\cite{photos}.

\section{Measurement method}
Using $e^+e^-$ annihilations at $\sqrt{s}=3.773$~GeV, we produce $D\bar{D}$
pairs with no additional hadrons.
Events where one $\bar{D}$ meson is fully reconstructed
are referred to as ``single-tag'' (ST) candidates.
A correct tag guarantees the presence of the other $D$ meson, and
we search for the hadronic decays $D^{0(+)}\to \eta\pi^{+}\pi^{-(0)}$
and $D^+\to\eta\eta\pi^+$ recoiling against a tagged $\bar{D}$ meson.
Events with both a tag and such a signal-mode candidate are referred
to as ``double-tag'' events (DT).
In this analysis, the tagged $\bar{D}^0$ mesons are reconstructed using three hadronic decay modes:
$K^+\pi^-$, $K^+\pi^-\pi^0$, and $K^+\pi^-\pi^-\pi^+$,
while the tagged $D^-$ mesons are reconstructed using six hadronic decay modes: $K^{+}\pi^{-}\pi^{-}$,
$K^0_{S}\pi^{-}$, $K^{+}\pi^{-}\pi^{-}\pi^{0}$, $K^0_{S}\pi^{-}\pi^{0}$, $K^0_{S}\pi^{+}\pi^{-}\pi^{-}$,
and $K^{+}K^{-}\pi^{-}$.
For a specific tag mode $i$, the yields of the tagged $\bar D$ mesons ($N_{\rm ST}^i$) and of the DT events ($N_{\rm DT}^i$) are
\begin{equation}
\label{Ni}
N_{\rm ST}^i= 2N_{D\bar{D}} {\mathcal B}_{\rm ST}^i \epsilon_{\rm ST}^i, \ \ \ \ N_{\rm DT}^i= 2N_{D\bar{D}} {\mathcal B}_{\rm ST}^i {\mathcal B}_{\rm sig} \epsilon_{\rm DT}^i {\mathcal B}_{\rm sub},
\end{equation}
where $N_{D\bar{D}}$ is the number of $D\bar{D}$ pairs, ${\mathcal B}_{\rm ST}^i$ and ${\mathcal B}_{\rm sig}$ are
the BFs of the $\bar{D}$ tag decay mode $i$ and the $D$ signal decay mode, $\epsilon_{\rm ST}^i$ is the efficiency for finding the tag candidate, and $\epsilon_{\rm DT}^i$ is the efficiency for simultaneously finding the tag $\bar{D}$ and the signal decay.   Finally, ${\mathcal B}_{\rm sub}$ is the appropriate BF product of $\eta\to\gamma\gamma$ and $\pi^0\to\gamma\gamma$ in the signal decay; i.e.,
${\mathcal B}_{\rm sub}$ is equal to ${\mathcal B}_{\eta\to\gamma\gamma}^2$, ${\mathcal B}_{\eta\to\gamma\gamma}{\mathcal B}_{\pi^0\to\gamma\gamma}$,
and ${\mathcal B}_{\eta\to\gamma\gamma}$ for $D^+\to\eta\eta\pi^+$, $D^+\to \eta\pi^+\pi^0$, and $D^0\to \eta\pi^+\pi^-$, respectively. Combining the above equation, the BF for the
signal decay is given by
\begin{equation}
\label{eq:br}
{\mathcal B}_{\rm sig} = \frac{N_{\rm DT}}{N_{\rm ST} \epsilon_{{\rm sig}}{\mathcal B}_{\rm sub}},
\end{equation}
where $N_{\rm ST}$ and $N_{\rm DT}$ are the total ST and DT yields and
$\epsilon_{{\rm sig}}$ is the average efficiency of reconstructing the signal decay (with a tag present), weighted by the measured yields of tag modes in data which is given by
\begin{equation}
\epsilon_{{\rm sig}}=\frac{\sum_i N^i_{{\rm ST}}  \epsilon^i_{{\rm DT}}/{\epsilon^i_{\rm ST}}}{\sum_i N^i_{{\rm ST}}}.
\end{equation}

\section{Event selection}

The event selection criteria used in this work are the same as those
used in Refs.~\cite{epjc76,cpc40,bes3-pimuv,liuke}.
All charged tracks are required to be within a polar-angle ($\theta$) range of
$|\rm{cos\theta}|<0.93$.
Except for those from $K^0_{S}$ decays, all tracks must originate
from an interaction region defined by
$V_{xy}<$ 1\,cm and $V_{z}<$ 10\,cm.
Here, $V_{xy(z)}$ is the distance of the closest approach
of the charged track to the interaction point perpendicular to (along) the beam.

Charged kaons and pions are identified
with the information of the TOF and the $dE/dx$ measured by the MDC.
Confidence levels
for pion and kaon hypotheses ($CL_{\pi}$ and $CL_{K}$) are calculated.
Kaon and pion candidates are required to satisfy $CL_{K}>CL_{\pi}$
and $CL_{\pi}>CL_{K}$, respectively.

The $K^0_S$ mesons are reconstructed in the decay
$K^0_S\to\pi^+\pi^-$.
Two oppositely charged tracks are required to
satisfy $V_{z}<$ 20\,cm, but without $V_{xy}$ and particle identification (PID) requirements.
The two tracks are constrained to originate from a common vertex, and their invariant mass is required to satisfy
$|M_{\pi^{+}\pi^{-}} - M_{K_{S}^{0}}|<$~12\,MeV$/c^{2}$, where
$M_{K_{S}^{0}}$ is the nominal mass~\cite{pdg2018}.
The vertex of the $K^0_S$ candidate is required to be more than
two standard deviations of the vertex resolution away from the interaction point.

The $\pi^0$ and $\eta$ mesons are reconstructed from their decay into two photons. Photon candidates are selected from the list of EMC showers. The shower time is required to be within 700\,ns of the event start time. The shower energy is required to be greater than 25 (50)\,MeV if the crystal with the maximum energy deposit in that cluster is in the barrel~(end cap) region~\cite{Ablikim:2009aa}. The opening angle between the candidate shower and
the nearest charged track must be greater than $10^{\circ}$.
Photon pairs with an invariant mass in the interval 0.115 $<M_{\gamma\gamma}<0.150$\,GeV/$c^2$ (0.515 $<M_{\gamma\gamma}<0.570$\,GeV/$c^2$)
are accepted as $\pi^0$ ($\eta$) candidates. To improve resolution, a one-constraint kinematic fit is imposed on each selected photon pair, in which the $\gamma\gamma$ invariant mass is constrained to the $\pi^{0}$ or $\eta$ nominal mass~\cite{pdg2018}.

In the selection of the tagged candidates of $\bar D^0\to K^+\pi^-$, backgrounds from cosmic rays and Bhabha events must be suppressed. First, the two charged tracks must have a TOF time difference less than 5 ns\,and they must not be consistent with being a muon pair or an electron-positron pair. Second,
there must be at least one EMC shower with an energy larger than 50 MeV or at least
one additional charged track detected in the MDC~\cite{deltakpi}.
Also, for the $D^0\to\eta\pi^+\pi^-$ candidate events, the invariant mass of the $\pi^+\pi^-$ combination is required to be outside the mass window of $|M_{\pi^+\pi^-} - M_{K_{S}^{0}}|<$~30\,MeV$/c^{2}$ to reject the backgrounds from the $D^0\to K^0_S\eta$ decays.

The tagged $\bar D$ (signal $D$) meson is identified by two variables,
the energy difference
\begin{equation}
\Delta E_{\rm tag\,(sig)} \equiv E_{\rm tag\,(sig)} - E_{\rm beam}
\label{eq:deltaE}
\end{equation}
and the beam-constrained mass
\begin{equation}
M_{\rm BC}^{\rm tag\,(sig)} \equiv \sqrt{E^{2}_{\rm beam}-|\vec{p}_{\rm tag\,(sig)}|^{2}},
\label{eq:mBC}
\end{equation}
where the superscript ${\rm tag\,(sig)}$ represents the tagged $\bar D$ candidate and signal $D$ candidate, $E_{\rm beam}$ is the beam energy,
and $\vec{p}_{\rm tag\,(sig)}$ and $E_{\rm tag\,(sig)}$ are the momentum and energy of the $\bar D\,(D)$ candidate in the rest frame of $e^+e^-$ system. For each tag (signal) mode, if there are multiple candidates in an event, only the one with the minimum $|\Delta E_{\rm tag\,(sig)}|$ is kept. The tag side is required to satisfy $\Delta E_{\rm tag}\in(-55,\, +40)$\,MeV for the modes containing a $\pi^0$ in the final state and $\Delta E_{\rm tag}\in(-25,\, +25)$\,MeV for the other modes. The signal side is required to satisfy $\Delta E_{\rm sig}\in(-42,\, +40)$, $(-68,\, +52)$, and $(-40,\, +38)$\,MeV for $D^+\to\eta\eta\pi^+$, $D^+\to \eta\pi^+\pi^0$, and $D^0\to \eta\pi^+\pi^-$, respectively.

\begin{figure*}[htp]
  \centering
\includegraphics[width=0.8\linewidth]{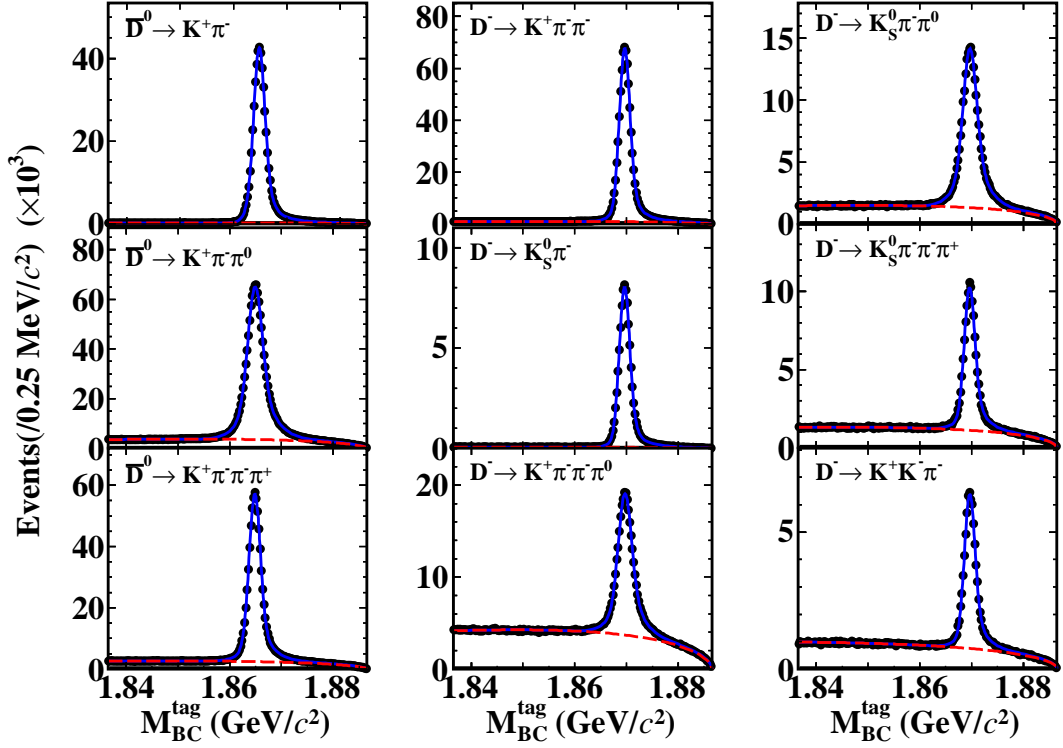}
  \caption{
Fits to the $M_{\rm BC}$ distributions of the $\bar D^0$ (left column) and $D^-$ (middle and right columns) tagging decay modes.
Data are shown as dots with error bars.
The blue solid and red dashed curves are the fit results
and the fitted backgrounds, respectively.
}
\label{fig:datafit_MassBC}
\end{figure*}

\section{SINGLE-TAG AND DOUBLE-TAG YIELDS}

The ST yields are obtained from maximum likelihood fits
to the $M_{\rm BC}^{\rm tag}$ distributions of the accepted tagged $\bar D$ candidates in data, as shown in Fig.~\ref{fig:datafit_MassBC}.
In the fits, the $\bar D$ signal is modeled by an MC-simulated shape via a RooHistPdf class in ROOT~\cite{RooHistPdf} convolved with
a double Gaussian function describing the resolution difference between data and MC simulation.
The combinatorial background shape is described by the ARGUS function~\cite{ARGUS}.
The ST yields and the ST efficiencies are summarized in Table~\ref{singletagN}. The total ST yields ($N_{\rm ST}$) are
$1558195\pm2113$ for $D^-$ and $2386575\pm1928$ for $\bar D^0$, where the uncertainties are statistical.
These yields are slightly different from those reported in Refs.~\cite{epjc76,cpc40,bes3-pimuv},
due to the lack of $M_{\rm BC}$ window requirements.

\begin{figure}[htp]
\centering
\includegraphics[width=0.75\linewidth]{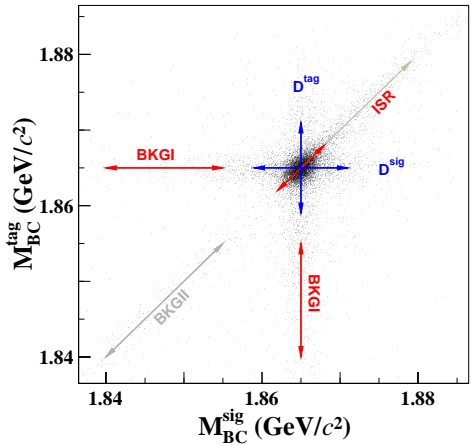}
\caption{
Illustration of the distributions of $M_{\rm BC}^{\rm tag}$
vs.~$M_{\rm BC}^{\rm sig}$ of the accepted DT hadronic $D\bar D$ candidate events.
}
\label{fig:mBC2D}
\end{figure}

\begin{table}[htp]
  \centering
    \caption{Summary of the ST yields ($N_{\rm ST}^i$)
    and the ST efficiencies ($\epsilon_{\rm ST}^i$) in data, where the uncertainties are statistical. The efficiencies do not include the BFs for $K_S^0\to \pi^+\pi^-$ and $\pi^0\to \gamma\gamma$.}
    \label{singletagN}
    \begin{tabular}{ccc}
      \hline
      \hline
        Tag mode & $N_{\rm ST}^i$&$\epsilon_{\rm ST}^i$ (\%)\\
	  \hline
	$K^{+}\pi^{-}$ &                    $ 527193\pm 761$&$65.60\pm0.09$\\
	$K^{+}\pi^{-}\pi^{0}$ &             $1138068\pm1373$&$37.69\pm0.04$\\
	$K^{+}\pi^{-}\pi^{-}\pi^{+}$ &      $ 721314\pm1120$&$38.98\pm0.06$\\
	        \hline
	$K^{+}\pi^{-}\pi^{-}$ &             $798935\pm1011$&$51.90\pm0.08$\\
	$K^{0}_{S}\pi^{-}$ &                $93308 \pm 329$&$51.80\pm0.17$\\
	$K^{+}\pi^{-}\pi^{-}\pi^{0}$ &      $258044\pm1036$&$26.92\pm0.09$\\
	$K^{0}_{S}\pi^{-}\pi^{0}$ &         $221792\pm1274$&$28.27\pm0.10$\\
	$K^{0}_{S}\pi^{-}\pi^{-}\pi^{+}$ &  $115532\pm 645$&$28.60\pm0.14$\\
	$K^{+}K^{-}\pi^{-}$ &               $70548 \pm 470$&$42.13\pm0.25$\\
		  \hline
		  \hline
    \end{tabular}
    \end{table}

Figure~\ref{fig:mBC2D} illustrates the distribution of $M_{\rm BC}^{\rm tag}$
vs.~$M_{\rm BC}^{\rm sig}$ for DT candidate events.
Signal events concentrate around $M_{\rm BC}^{\rm tag} = M_{\rm BC}^{\rm sig} = M_{D}$,
where $M_{D}$ is the nominal $D$ mass~\cite{pdg2018}.
Background events are divided into three categories.
The first one, BKGI, is from events with correctly reconstructed $D$ ($\bar D$) and incorrectly
reconstructed $\bar D$ ($D$), which are spread along the lines where
either $M_{\rm BC}^{\rm tag}$ or $M_{\rm BC}^{\rm sig}$ equals $M_{D}$.
The second one, BKGII, is from events spread along the diagonal,
which are mainly from the $e^+e^- \to q\bar q$ processes.
The third one, BKGIII, comes from events with both $D$ and $\bar D$ reconstructed incorrectly which spread out the full plot. To extract the
DT yield in data, a two-dimensional (2D) unbinned maximum likelihood fit~\cite{cleo-2Dfit} on this distribution is performed. In the fit, the probability density functions (PDFs) of the
four components mentioned above are constructed as
\begin{widetext}
\begin{itemize}
\item signal: $a(M_{\rm BC}^{\rm sig},M_{\rm BC}^{\rm tag})$,
\item BKGI:
$b(M_{\rm BC}^{\rm {sig}})\cdot c(M_{\rm BC}^{\rm {tag}};E_{\rm beam},\xi_{M_{\rm BC}^{\rm tag}},\frac{1}{2})$+$b(M_{\rm BC}^{\rm {tag}})\cdot c(M_{\rm BC}^{\rm {sig}};E_{\rm beam},\xi_{M_{\rm BC}^{\rm sig}},\frac{1}{2})$,
\item BKGII:
 $c((M_{\rm BC}^{\rm sig}+M_{\rm BC}^{\rm tag})/\sqrt{2};\sqrt{2}E_{\rm beam},\xi,\frac{1}{2})\cdot
  (Gg((M_{\rm BC}^{\rm sig}-M_{\rm BC}^{\rm tag})/\sqrt{2};0,\sigma_{1}) +
  (1-G)g((M_{\rm BC}^{\rm sig}-M_{\rm BC}^{\rm tag})/\sqrt{2};0,\sigma_{2}))$,
\item BKGIII:
$c(M_{\rm BC}^{\rm sig};E_{\rm beam},\xi_{M_{\rm BC}^{\rm sig}},\frac{1}{2}) \cdot c(M_{\rm BC}^{\rm tag};E_{\rm beam},\xi_{M_{\rm BC}^{\rm tag}},\frac{1}{2})$,
\end{itemize}
\end{widetext}

\noindent where $g(x;0,\sigma)$ denotes a Gaussian function with mean of zero and standard deviation of $\sigma$,
 $c(x;E_{\rm beam},\xi,\frac{1}{2})$ is an ARGUS function defined as
$Ax(1 - \frac {x^2}{E_{\rm beam}^2})^{\frac{1}{2}} \cdot e^{\xi(1-\frac {x^2}{E_{\rm beam}^2})}$.
Here, $A$ is a normalization constant (independent for the ARGUS functions in the $M^{\rm sig}_{\rm BC}$ and $M^{\rm tag}_{\rm BC}$ directions), $E_{\rm beam}$ is the end point which is fixed at 1.8865 GeV, and $G$ is the fraction of two Gaussians.
The PDFs of signal $a(M_{\rm BC}^{\rm sig},M_{\rm BC}^{\rm tag})$, $b(M_{\rm BC}^{\rm sig})$, and $b(M_{\rm BC}^{\rm tag})$ are described by the corresponding MC-simulated shapes, with the kernel-estimation method~\cite{keyspdf} via a RooNDKeysPdf class in ROOT~\cite{keyspdf2}. Other parameters are left free.

There are some peaking backgrounds in $M_{\rm BC}^{\rm tag}$ vs.~$M_{\rm BC}^{\rm sig}$ distribution to consider. For the decay $D^+\to\eta\eta\pi^+$, the peaking backgrounds are from a correct tag with an incorrect signal ($D^+\to\pi^+\pi^0\pi^0$).
For the decay $D^+\to\eta\pi^0\pi^+$, the peaking backgrounds are from a correct tag with an incorrect signal [$D^+\to K_L^0(K^0_S)\pi^+\pi^0$, $K_S^0\to \pi^0\pi^0$, or $D^+\to\pi^+\pi^0\pi^0$]. For these peaking backgrounds, the shapes are modeled based on MC simulation and the normalizations are fixed according to the corresponding BFs in PDG~\cite{pdg2018}.

\begin{figure}[htp]
  \centering
\includegraphics[width=1.0\linewidth]{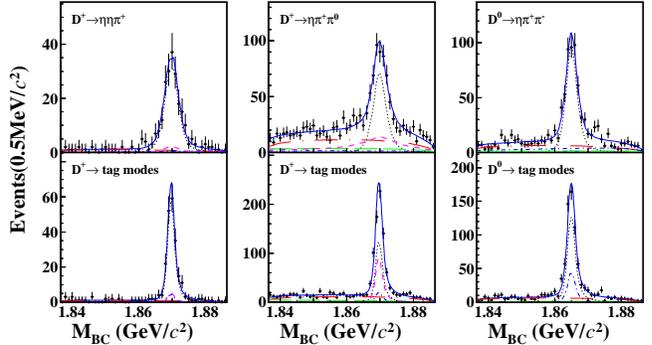}
  \caption{
The projections on $M^{\rm tag}_{\rm BC}$ (bottom) and
$M^{\rm sig}_{\rm BC}$ (top) of the 2D fits to the DT candidate
events for $D^+\to\eta\eta\pi^+$ (left), $D^+\to\eta\pi^0\pi^+$ (middle), and $D^0\to\eta\pi^+\pi^-$ (right).
Data are shown as dots with error bars.
The blue solid, black dotted, blue dot-dashed, red dot-long-dashed,
green long-dashed, and pink dashed curves denote the overall fit results,
signal, BKGI, BKGII, BKGIII, and peaking background components (see the text),
respectively.
}
\label{fig:2Dfit}
\end{figure}

Figure~\ref{fig:2Dfit} shows the $M_{\rm BC}^{\rm tag}$ and $M_{\rm BC}^{\rm sig}$ projections of the 2D fits to data.
From these 2D fits, we obtain the DT yields for individual signal decays ($N_{\rm DT}$) in the fitted $M_{\rm BC}^{\rm tag\,(sig)}$ region (1.8365, 1.8865) GeV$/c^2$, as shown in the second column of Table \ref{tab:signal}.
For each signal decay mode, the statistical significance is calculated according to
$\sqrt{-2{\rm ln}(\mathcal L_0/\mathcal L_{\rm max})} $,
where $\mathcal L_{\rm max}$ is the maximal likelihood of the nominal fit and $\mathcal L_0$ is the likelihood of the corresponding fit without the signal component.
The statistical significance for the three signal decays are all found to be greater than  $10\sigma$.

\section{BRANCHING FRACTIONS}

\begin{figure*}[htp]
  \centering
  \includegraphics[width=0.95\linewidth]{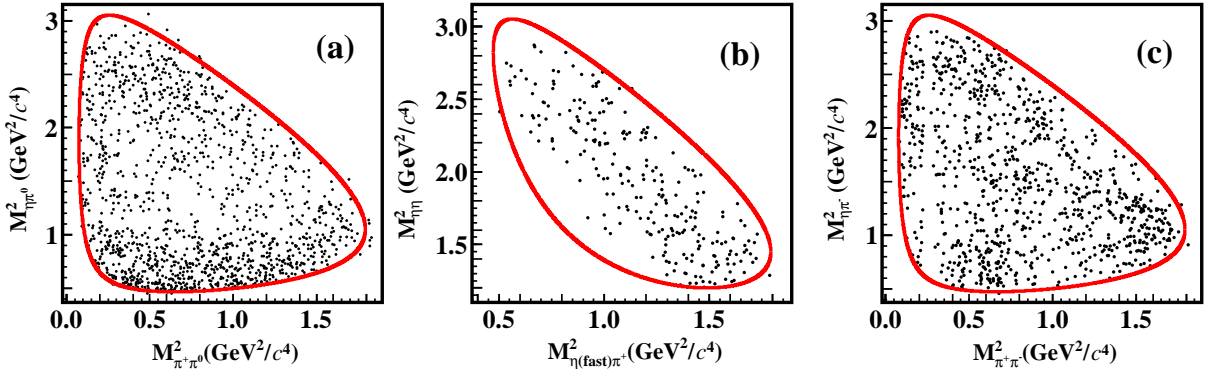}
    \caption{
	   Dalitz plots of (a) $M_{\pi^+\pi^0}^2$ vs.~$M_{\eta\pi^0}^2$ for $D^+\to\eta\pi^+\pi^0$, (b) $M^2_{\eta(\rm fast)\pi^+}$ vs.~$M^2_{\eta\eta}$ for $D^+\to\eta\eta\pi^+$, and (c) $M^2_{\pi^+\pi^-}$ vs.~$M^2_{\eta\pi^-}$ for $D^0\to\eta\pi^+\pi^-$ in data. In these figures, all selection criteria have been imposed and the $M_{\rm BC}^{\rm tag(sig)}$ is required to be within 6 MeV/$c^2$ of the nominal $D$ mass~\cite{pdg2018}. The red curves show the kinematically allowed regions.
	    }
	    \label{com_2D}
\end{figure*}

To ensure the reliability of signal efficiency, we have examined the $M_{\eta P}$, $M_{\eta \pi^+}$,
and $M_{P\pi^+}$ distributions of $D\to \eta P \pi^+$ candidate events after requiring $|M_{\rm BC}^{\rm sig}-M_D|<0.006$ GeV/$c^2$.
Here, $P$ denotes the daughter particles of $\eta$, $\pi^0$, and $\pi^-$ for $D^+\to \eta\eta\pi^+$, $D^+\to \eta\pi^+\pi^0$,
and $D^0\to \eta\pi^+\pi^-$ decays, respectively.
Figure~\ref{com_2D} shows the Dalitz plots of three signal decay modes in data, and there are
no significant $\rho^{0,\pm}$ and $a_0(980)^{0,\pm}$ signals in these Dalitz plots.
However, due to some possible resonances, the phase-space MC distributions of $M_{\eta P}$, $M_{\eta \pi^+}$, and $M_{P\pi^+}$ do not agree well with the data distributions.  To solve this problem, the MC generator is modified to produce the correct invariant mass distributions according to the Dalitz plots in data. In the Dalitz plot, the background component is modeled by the inclusive MC simulation, while the signal components generated according to an efficiency-corrected MC simulation. These modified MC samples are in good agreement with the data distributions and are therefore used to determine the averaged efficiencies of the signal decays ($\epsilon_{\rm sig}$), which are summarized in Table~\ref{tab:signal}.

\begin{table*}[htp]
\centering
\caption{\label{tab:signal}
\small
The DT yields in data ($N_{\rm DT}$),  signal efficiencies ($\epsilon_{\rm sig}$), obtained BFs ($\mathcal B_{\rm sig}$), and the corresponding BFs ($\mathcal B_{\rm CLEO}$) measured by CLEO~\cite{cleoeta}.
The efficiencies do not include the BFs of $\eta \to\gamma\gamma$ and $\pi^0\to \gamma\gamma$.
The uncertainties in $N_{\rm DT}$ and $\epsilon_{\rm sig}$ are statistical. The first and second uncertainties of
$\mathcal B_{\rm sig}$ and $\mathcal B_{\rm CLEO}$ are statistical and systematic, respectively.
}
\begin{tabular}{lcccc}
\hline\hline
Decay mode             & $N_{\rm DT}$ & $\epsilon_{\rm sig}$\,(\%)& $\mathcal B_{\rm sig}$\,($\times 10^{-3}$)& $\mathcal B_{\rm CLEO}$\,($\times 10^{-3}$)\\ \hline
$D^+\to \eta\eta\pi^+$ &$179\pm15$&$24.96\pm0.12$&$2.96 \pm 0.24 \pm 0.10$&N/A \\
$D^+\to \eta\pi^+\pi^0$&$381\pm26$&$28.11\pm0.13$&$2.23 \pm 0.15 \pm 0.10$&$1.38 \pm 0.31\pm0.16$ \\
$D^0\to \eta\pi^+\pi^-$&$450\pm25$&$39.98\pm0.17$&$1.20 \pm 0.07 \pm 0.04$&$1.09 \pm 0.13\pm0.09$ \\
\hline\hline
\end{tabular}
\end{table*}

The absolute BFs of the signal decays obtained according to Eq.~(\ref{eq:br}), are summarized in Table~\ref{tab:signal}.

The BFs of $D\to f$ and ${\bar{D}\to \overline{f}}$ are also measured separately for each final state $f$.
The asymmetry of the BFs of the $D$ and $\bar D$ decays is determined by
$A_{CP}=\frac{{\mathcal B}(D\to f)-{\mathcal B}(\bar{D}\to {\overline{f}})}{{\mathcal B}(D\to f)+{\mathcal B}(\bar{D}\to {\overline{f}})}$.
The ST yields ($N_{\rm ST}$), the DT yields ($N_{\rm DT}$), the signal efficiencies ($\epsilon_{\rm sig}$), and the obtained BFs ($\mathcal B_{\rm sig}$) for $D$ and $\bar D$ decays, as well as the determined $A_{CP}$ values are summarized in Table~\ref{CP_result}.

\begin{table}[htp]
\centering
\caption{
\small
Summary of the ST yields ($N_{\rm ST}$), the signal yields ($N_{\rm DT}$), and the signal efficiencies ($\epsilon_{\rm sig}$) used to determine the BFs ($\mathcal B_{\rm sig}$) and $CP$ asymmetries ($A_{CP}$) for $D\to \rm sig$ and $\bar {D}\to \overline{\rm sig}$. For $A_{CP}$, the first and second uncertainties are statistical and systematic, respectively. The uncertainties for other values are only statistical.
}
\label{CP_result}
\centering
\begin{tabular}{lccc}
  \hline\hline
      &$D^-\to\eta\eta\pi^-$&$D^-\to\eta\pi^-\pi^0$&$\bar{D}^0\to\eta\pi^+\pi^-$\\
       \hline
      $N_{\rm ST}$                                       &777280$\pm$1466   &777280$\pm$1466   &1188894$\pm$1329\\
      $N_{\rm DT}$                                       &81$\pm$10         &202$\pm$19        &245$\pm$18\\
      $\epsilon_{\rm sig}$ (\%)                          &25.08$\pm$0.17    &28.13$\pm$0.18    &39.94$\pm$0.24\\
      $\mathcal B~(\times 10^{-3})$                      &2.69$\pm$0.34     &2.37$\pm$0.22     &1.31$\pm$0.09\\
     \hline
      &$D^+\to\eta\eta\pi^+$&$D^+\to\eta\pi^+\pi^0$&$D^0\to\eta\pi^+\pi^-$\\
     \hline
      $N_{\rm ST}$                                       &782704$\pm$1491   &782704$\pm$1491   &1197025$\pm$1374\\
      $N_{\rm DT}$                                       &96$\pm$11         &182$\pm$17        &204$\pm$17\\
      $\epsilon_{\rm sig}(\%)$                           &25.03$\pm$0.17    &28.21$\pm$0.18    &40.07$\pm$0.23\\
      $\mathcal B~(\times 10^{-3})$                      &3.16$\pm$0.35     &2.11$\pm$0.20     &1.08$\pm$0.09\\
        \hline
      $A_{CP}$ (\%)                                      &8.0$\pm$8.3$\pm$1.9&$-$5.8$\pm$6.6$\pm$1.8&$-$9.6$\pm$5.4$\pm$1.8\\
        \hline\hline
\end{tabular}
\end{table}

\section{Systematic uncertainties}
\label{sec:sys}

With the DT method, most of uncertainties related to the tagged $\bar D$ are canceled.
A summary of the systematic uncertainties in the BF measurements is given in
Table~\ref{tab:relsysuncertainties} and is discussed below.

\begin{itemize}
\item
{\bf ST yields:}
The uncertainties in the total ST yields come from the fits to the $M_{\rm BC}$ spectra of the tagged
$\bar D^0$ and $D^-$ candidates. They have been previously estimated to be
0.5\% for both neutral and charged $D$ in Refs.~\cite{epjc76,cpc40,bes3-pimuv}.

\item
{\bf  Tracking (PID) of $\pi^\pm$}:
The tracking (PID) efficiencies of $\pi^\pm$
are investigated with DT $D\bar D$ hadronic events by using a
partial reconstruction technique.
The systematic uncertainty for each charged particle due to tracking (PID)
is estimated to be 0.5\% (0.5\%).

\item
{\bf $\pi^0(\eta)$ reconstruction}:
The efficiency of $\pi^0$ reconstruction is studied with
the DT $D\bar D$ hadronic decays $D^0\to K^-\pi^+$, $K^-\pi^+\pi^+\pi^-$ vs.~$\bar D^0\to K^+\pi^-\pi^0$, $K^0_S\pi^0$~\cite{epjc76,cpc40}.
A small data-MC difference in the $\pi^0$ reconstruction efficiency is found.
The momentum weighted data-MC difference in $\pi^0$ reconstruction efficiencies is found to be
$(-0.5\pm 1.0)\%$, where the uncertainty is statistical. After correcting the MC efficiencies by the
momentum weighted data-MC difference in $\pi^0$ reconstruction efficiency, the systematic uncertainty
due to $\pi^0$ reconstruction is assigned as 1.0\% per $\pi^0$. The systematic uncertainty due to $\eta$
reconstruction is assumed to be the same as $\pi^0$ reconstruction and fully correlated.

\item
{\bf 2D yield fits}:
The systematic uncertainty due to the 2D fits of the $M^{\rm tag}_{\rm BC}$ vs.~$M^{\rm sig}_{\rm BC}$ distributions is evaluated by repeating
the measurements with an alternative fit range of $(1.8300,1.8865)$\,GeV/$c^2$,
an alternative signal shape with different MC matching requirements,
alternative end points of the ARGUS function, $E_{\rm beam}\pm0.2$\,MeV/$c^2$,
and with the quoted BFs of peaking backgrounds varied by $\pm1\sigma$.
The total systematic uncertainties are assigned based on the changes of the BFs from each of these sources summed in quadrature, yielding 1.0\%, 2.1\%, and 0.8\% for
$D^+\to\eta\eta\pi^+$, $D^+\to\eta\pi^+\pi^0$, and $D^0\to \eta\pi^+\pi^-$,
respectively.

\item
{\bf $\Delta E_{\rm sig}$ requirement}:
The systematic uncertainties due to the $\Delta E_{\rm sig}$ requirement are assigned by comparing the DT efficiencies with and without smearing by
the data-MC difference of the $\Delta E_{\rm sig}$ resolution for the signal MC events.
Here, the $\Delta E_{\rm sig}$ resolution differences
are obtained by using larger DT samples of $D^0\to K^-\pi^+\eta$, $D^0\to K^0_S\eta$, and $D^+\to \pi^+\pi^0\pi^0$ with the same tags.
The maximum change of the DT efficiency is taken to be the systematic uncertainties,
which is 0.3\% for
$D^+\to\eta\eta\pi^+$, $D^+\to\eta\pi^+\pi^0$, and $D^0\to \eta\pi^+\pi^-$.

\item
{\bf Modified MC generator}:
The differences between the signal efficiencies obtained with the PHSP MC and modified MC models are only 1.5\%, 1.2\%, and 0.5\% for $D^+\to\eta\eta\pi^+$, $D^+\to\eta\pi^0\pi^+$, and $D^0\to\eta\pi^-\pi^+$, respectively. No large uncertainty in the modified MC generator is foreseen.
Since the systematic uncertainties arising from the $\pi^\pm$ tracking and PID efficiencies as well as the $\eta$ and $\pi^0$ reconstruction efficiencies have been taken into account as independent sources, the systematic uncertainty in the modified MC generator is studied
with an alternative input Dalitz plot obtained by varying the MC-simulated background sizes.
The largest changes of the detection efficiencies, 2.1\%, 3.3\%, and 1.8\% for
$D^+\to\eta\eta\pi^+$, $D^+\to\eta\pi^+\pi^0$, and $D^0\to \eta\pi^+\pi^-$
are taken as the systematic uncertainties.

\item
{\bf MC statistics}:
The uncertainties due to the limited MC statistics are 0.5\%, 0.5\%, and 0.4\%
$D^+\to\eta\eta\pi^+$, $D^+\to\eta\pi^+\pi^0$, and $D^0\to \eta\pi^+\pi^-$, respectively.

\item
{\bf $K_S^0$ rejection}:
The efficiency uncertainty from $K_S^0$ rejection is estimated by using an alternative rejection window of $\pm40$ MeV/$c^2$ around the $K^0_S$ nominal mass.
The change in the BF, 1.4\%, is assigned as the systematic uncertainty for $D^0\to \eta\pi^+\pi^-$.

\item
{\bf Quoted BFs}:
The uncertainties of the quoted BFs of $\eta\to \gamma\gamma$ and
$\pi^0\to \gamma\gamma$~\cite{pdg2018}
are 0.5\% and 0.03\%, respectively.
The associated systematic uncertainties are 1.0\%, 0.5\%, and 0.5\% for
$D^+\to\eta\eta\pi^+$, $D^+\to\eta\pi^+\pi^0$, and $D^0\to \eta\pi^+\pi^-$, respectively.

\item
{\bf \boldmath Asymmetry of $CP\pm$ components}:
The measurement of the BF of $D^0\to \eta\pi^+\pi^-$ is affected by $CP\pm$ eigenstate components in the $D^0\to \eta\pi^+\pi^-$ decay. The asymmetry of $CP+$ and $CP-$ components in this decay is
examined by the $CP+$ tag of $D^0 \to K^+K^-$ and the $CP-$ tag of $D^0\to K^0_S\pi^0$.
Combined with the strong-phase factors of the flavor tags $\bar {D}^0\to K^-\pi^+$, $\bar {D}^0\to K^-\pi^+\pi^0$, and $\bar {D}^0\to K^-\pi^+\pi^+\pi^-$~\cite{pdg2018,spf,spf1},
the impact on the BF of $D^0\to \eta\pi^+\pi^-$ is found to be (1.0$\pm$0.9)\% with the same method described in Ref.~\cite{qusq}.
After correcting the BF of $D^0\to \eta\pi^+\pi^-$ by this factor, 0.9\% is assigned as an associated uncertainty.

\end{itemize}

The total systematic uncertainty obtained by adding the above
contributions in quadrature is 3.4\%, 4.5\%, and 3.2\% for
$D^+\to \eta\eta\pi^+$, $D^+\to \eta\pi^+\pi^0$, and $D^0\to \eta\pi^+\pi^-$,
respectively.

\begin{table}[htp]
\centering
\caption{\label{tab:relsysuncertainties}
Relative systematic uncertainties (in \%) in the BF measurements.}
\begin{small}
\begin{tabular}{lccc}
\hline\hline
Source & $\eta\eta\pi^+$ &$\eta\pi^+\pi^0$ &$\eta\pi^+\pi^-$\\
\hline
ST yield                    & 0.5 & 0.5 & 0.5 \\
Tracking of $\pi^\pm$       & 0.5 & 0.5 & 1.0 \\
PID of $\pi^\pm$            & 0.5 & 0.5 & 1.0 \\
$\pi^0\,(\eta)$ reconstruction& 2.0 & 2.0 & 1.0 \\
2D fit on $M^{\rm tag}_{\rm BC}$ vs.~$M^{\rm sig}_{\rm BC}$& 1.0 & 2.1 & 0.8 \\
$\Delta E_{\rm sig}$ requirement& 0.3 & 0.3 & 0.3 \\
Modified MC generator       & 2.1 & 3.3 & 1.8 \\
MC statistics               & 0.5 & 0.5 & 0.4 \\
$K_S^0$ rejection           &  -- & -- & 1.4 \\
Quoted BFs                  & 1.0 & 0.5 & 0.5 \\
Asymmetry of $CP\pm$ components& --  & --  & 0.9 \\
\hline
Total                       & 3.4 & 4.5 & 3.2 \\
\hline\hline
\end{tabular}
\end{small}
\end{table}

In the determinations of $A_{CP}$, the uncertainties of $\pi^0$ and $\eta$ reconstruction, quoted BFs, MC modeling, measurement method for each decay,
$\pi^+\pi^-$ tracking and PID as well as strong phase for $D^0(\bar{D}^0)\to \eta\pi^+\pi^-$ are assumed to cancel, while for $D^{+/-}\to \eta \pi^{+/-}\pi^0$ and $\eta\eta\pi^{+/-}$ decays, the uncertainties of $\pi^{+/-}$ tracking and PID are assumed to be uncanceled.
The remaining systematic uncertainties have been estimated separately with the same methods mentioned above.
With current statistics, no evidence of $CP$ violation is found.

\section{Conclusions}

With a data sample corresponding to an integrated luminosity of $2.93\,\rm fb^{-1}$
taken at $\sqrt{s}=3.773$\,GeV with the BESIII detector, we measure
the absolute BFs of the singly Cabibbo-suppressed
decays $D^+\to\eta\eta\pi^+$, $D^+\to\eta\pi^+\pi^0$, and $D^0\to\eta\pi^+\pi^-$.
The BF of $D^+\to\eta\eta\pi^+$ is measured for the first time.
The BFs of $D^+\to\eta\pi^+\pi^0$ and $D^0\to\eta\pi^+\pi^-$ are consistent with the CLEO-c's results~\cite{cleoeta} within $2.2\sigma$ and $0.6\sigma$, respectively.
The asymmetries of the BFs of $D$ and $\bar D$ decays in the three channels have also been examined, and no evidence of $CP$ violation is found.
In the near future, amplitude analyses of these three decays with larger data samples at BESIII and Belle II will offer the opportunity to explore two-body decays $D\to \rho\eta$, $a_0(980)\pi$, and $a_0(980)\eta$.

\section{Acknowledgements}

The BESIII Collaboration thanks the staff of BEPCII and the IHEP computing center for their strong support. This work is supported in part by National Key Basic Research Program of China under Contract No. 2015CB856700; National Natural Science Foundation of China (NSFC) under Contracts Nos. 11775230, 11475123, 11625523, 11635010, 11735014, 11822506, 11835012, 11935015, 11935016, 11935018, 11961141012; the Chinese Academy of Sciences (CAS) Large-Scale Scientific Facility Program; Joint Large-Scale Scientific Facility Funds of the NSFC and CAS under Contracts Nos. U1532101, U1732263, U1832207, U1932102; CAS Key Research Program of Frontier Sciences under Contracts Nos. QYZDJ-SSW-SLH003, QYZDJ-SSW-SLH040; 100 Talents Program of CAS; INPAC and Shanghai Key Laboratory for Particle Physics and Cosmology; ERC under Contract No. 758462; German Research Foundation DFG under Contracts Nos. Collaborative Research Center CRC 1044, FOR 2359; Istituto Nazionale di Fisica Nucleare, Italy; Ministry of Development of Turkey under Contract No. DPT2006K-120470; National Science and Technology fund; STFC (United Kingdom); The Knut and Alice Wallenberg Foundation (Sweden) under Contract No. 2016.0157; The Royal Society, UK under Contracts Nos. DH140054, DH160214; The Swedish Research Council; U. S. Department of Energy under Contracts Nos. DE-FG02-05ER41374, DE-SC-0012069.

\end{document}